\documentclass[draft]{aipproc}
\layoutstyle{6x9}

\newcommand{\bbox}[1]{\kern 1pt #1\kern -8pt #1\kern 1pt}

\begin{document}
\begin{flushright}
FERMILAB-Conf-01/368-T \\
hep-ph/0111376
\end{flushright}

\title[\sl 9th International Symposium on Heavy Flavor Physics]%
{Lattice QCD and the Unitarity Triangle}


\author{Andreas S. Kronfeld}{
  address={Fermi National Accelerator Laboratory, P.O. Box 500,
  Batavia, Illinois, USA},
  email={ask@fnal.gov},
  thanks={Fermilab is operated by Universities Research Association Inc.,
under contract with the U.S.\ Department of Energy.}
}

\copyrightyear{2001}

\begin{abstract}
Theoretical and computational advances in lattice calculations
are reviewed, with focus on examples relevant to the unitarity triangle
of the CKM matrix.
Recent progress in semi-leptonic form factors for $B\to\pi l\nu$ and
$B\to D^*l\nu$, as well as the parameter $\xi$ in $B^0$-$\bar{B}^0$ 
mixing, are highlighted.
\end{abstract}

\date{\rm September 10--13, 2001}

\maketitle

\section{Introduction}

\begin{figure}[!b]
\setlength{\unitlength}{0.5pt}
\begin{picture}(336,125)(-60,-30)

\thicklines
\put(0,0){\vector(1,4){28}}
\put(70,-24){$A=V_{cd}V_{cb}^*$}
\put(252,0){\vector(-1,0){252}}
\put(-120,56){$B=V_{ud}V_{ub}^*$}
\put(12,12){$\gamma$}

\thinlines
\put(28,112){\line(2,-1){224}}
\end{picture}
\hfill
\begin{picture}(336,125)(-120,-30)
\thicklines
\put(28,112){\vector(2,-1){224}}
\put(160,66){$C=V_{td}V_{tb}^*$}
\put(32,84){$\alpha$}
\put(180,9){$\beta$}

\thinlines
\put(0,0){\line(1,4){28}}
\put(252,0){\line(-1,0){252}}
\end{picture}
\caption{Unitarity triangles: on the left is the ``tree'' triangle;
on the right, the ``mixing'' triangle.}
\label{fig:ut}
\end{figure}
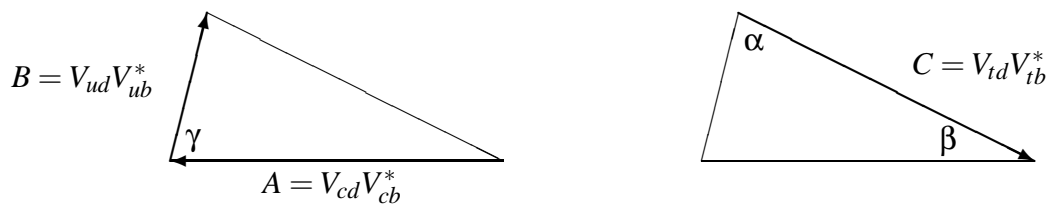

To test the CKM picture of $CP$ and flavor violation, a combination of 
theory and experiment is needed.
A vivid way to summarize the need for redundant information 
is the unitarity triangle (UT), sketched in Fig.~\ref{fig:ut}.
It depicts two triangles, 
$B\gamma A$, which can (in principle) be determined from tree 
processes (e.g., semi-leptonic decays), and $\alpha C\beta$, 
in which the amplitude for $B^0$-$\bar{B}^0$ mixing is 
involved~\cite{Kronfeld:2000id}.
The ``tree'' triangle can, for the sake of argument, be taken as a 
measurement of the CKM matrix.
Then, the ``mixing'' triangle tests whether new physics must be 
invoked to explain $B^0$-$\bar{B}^0$ mixing.

To obtain the sides $A$, $B$, and $C$ of these triangles one must 
calculate hadronic properties from first principles of~QCD.
Here ``calculate'' implies that a reliable error estimate is given:
one that includes all sources of uncertainty.
Lattice gauge theory is well suited to the task:
semi-leptonic form factors and the mixing matrix elements are 
conceptually straightforward.
Indeed, according to Martin Beneke~\cite{Beneke:2001bl},
``[the] standard UT fit is now entirely in the hands of lattice QCD
(up to, perhaps, $|V_{ub}|$).''

Until recently, lattice QCD has been burdened by something called 
the quenched approximation (explained below).
A~bit of good news is that the available computer power is now 
sufficient to get rid of this
approximation~\cite{AliKhan:2000eg,AliKhan:2001eg,Bernard:2001wy}.
Another bit of good news is that several lattice groups have used the 
quenched approximation in the spirit of a blind analysis: although 
quenching could change the central value, one can analyze all other 
uncertainties as if the underlying numerical data were real~QCD.
This exercise has left us with several quantities, including those 
needed in the UT fits, with full error analyses, apart from quenching.
In the next few years, we should have full QCD calculations with
a complete assessment of the uncertainties, and thereafter the
uncertainties can be incrementally reduced.

Because it is important to understand the theoretical uncertainties,
this talk will start by sketching where they arise.
Most reviews focus too much on central values, so, even when turning 
to numerical results, the focus here is on the error bars.

The paper ends with a short swim through the ``Octopus's Garden'',
that is, some recent work by Nathan Isgur, who passed a way a few
weeks before Heavy Flavors~9 convened.
The symposium, and this paper, are dedicated to his memory.

\section{Uncertainties in Lattice Calculations}

Lattice QCD calculates matrix elements by integrating the functional
integral, using a Monte Carlo with importance sampling.
The Monte Carlo leads to (correlated) statistical error bars.
This part of the method is well understood for quenched QCD and, these 
days, rarely leads to controversy.
When conflicting results arise, they originate in different
treatments of the systematics.
The consumer probably does not need to know how the Monte Carlo works,
but should develop an intuition of how the systematics work.

The main tool for controlling systematics is effective field theory.
In this talk, we are concerned mostly with three classes of effects:
those connected with the lattice spacing~$a$,
the heavy quark mass(es)~$m_Q$, and the light quark mass(es)~$m_q$.
Inside the computer there is a hierarchy of scales
\begin{equation}
	m_q \ll \Lambda \ll m_Q,\;\pi/a,
\end{equation}
although in practice $\pi/a$ is a only several GeV, and the ``light''
quarks are never as light as the up and down quarks.
The QCD scale $\Lambda$ gauges the size of power corrections;
experience suggests it is 700~MeV, give or take a factor of~$\sqrt{2}$.
With familiar techniques of effective field theory, lattice theorists
can control the extrapolation of artificial, numerical data to the
real world, provided the data start ``near'' enough.
Similarly, non-experts usually have an intuition of how effective field
theories work, so they can check, on the back of the envelope, whether
systematic errors have been treated sensibly.

\begin{figure}[b]
	\includegraphics[width=0.44\textwidth]{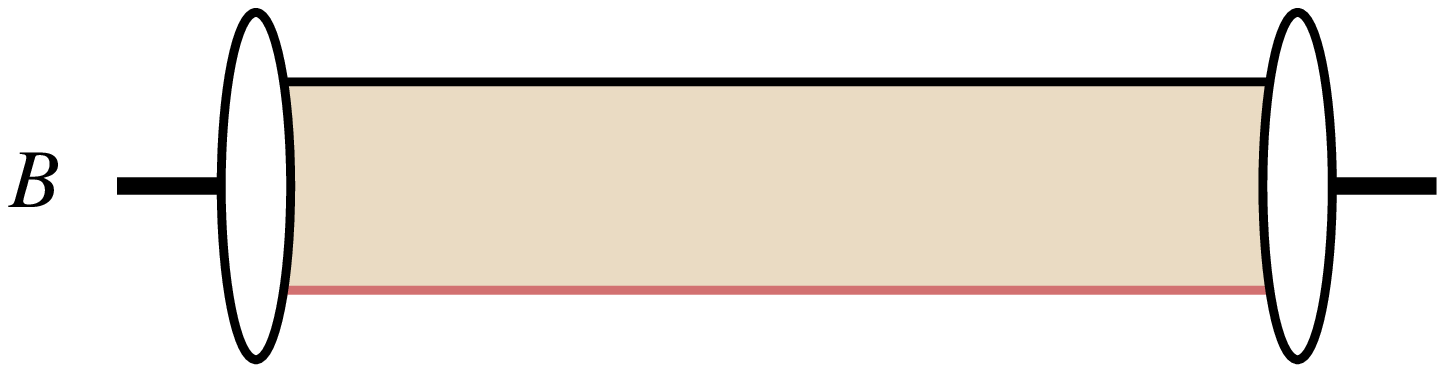}\hspace{0.06\textwidth}
	\includegraphics[width=0.44\textwidth]{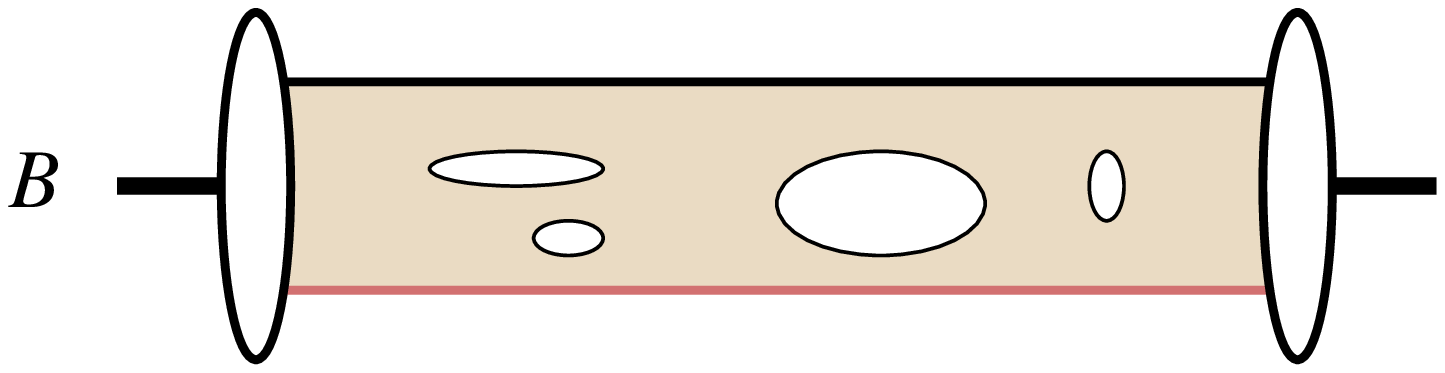}
	\caption{Quenched approximation: processes on the left are
	incorporated ``exactly'', whereas processes on the right are not
	computed, but modeled with a shift in the bare couplings.}
    \label{fig:quench}
\end{figure}

The notable exception to the rule of effective field theory is the
valence, or quenched, approximation.
Consider the pictures in Fig.~\ref{fig:quench}.
The one on the left depicts a meson made of a valence quark and 
anti-quark, bound by a shmear of gluons.
The gluons can also create virtual quark-antiquark pairs, leading to the 
picture on the right.
These are computationally very costly.
The quenched approximation omits them but compensates the omission with
a shift in the bare couplings.
This is analogous to a dielectric, where $g_0^2\to g_0^2\epsilon$.
Quenching retains many effects, such as retarded gauge potentials,
that are omitted in, say, the quark model.
It is also the first term in a systematic expansion~\cite{Sexton:1997ud}.

\begin{figure}[b]
	\includegraphics[width=0.3\textwidth]{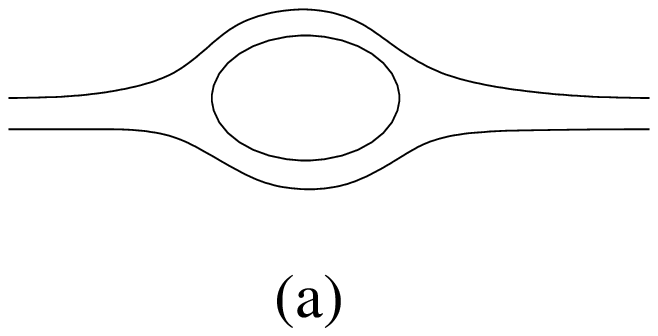}\hspace{.2in}
	\includegraphics[width=0.3\textwidth]{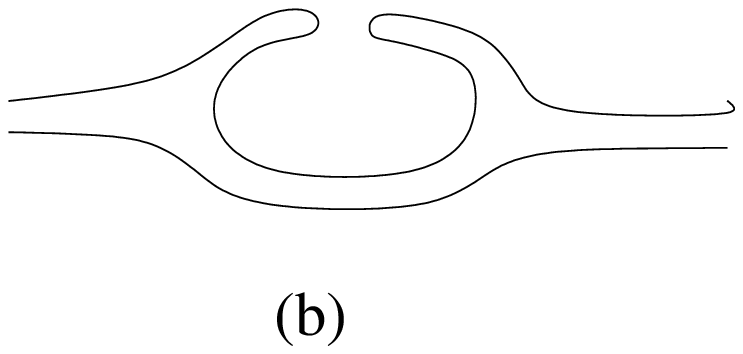}\hspace{.2in}
	\includegraphics[width=0.3\textwidth]{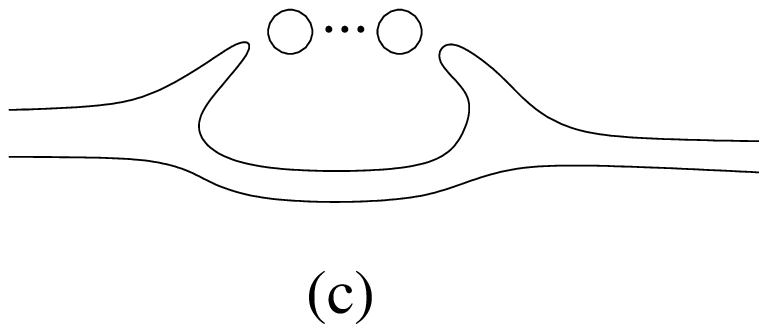}
	\caption{Quark line configurations that lead to meson loops.
	Adapted from Ref.~\cite{Bardeen:2001jm}.}
	\label{fig:lines}
\end{figure}

The quenched approximation can fall short of reproducing chiral
logarithms of the form $\ln(\Lambda^2/m_q^2)$
\cite{Morel:1987xk,Sharpe:1990me,Bernard:1992mk,Bernard:1994sv}.
Figure~\ref{fig:lines} shows some quark-line configurations that 
generate, at the hadronic level, meson loops.
The quenched approximation includes~(b), but not (a) or (c).
As a consequence, some pion loops are omitted, and $\eta'$ loops are
mistreated.
A better situation is a ``partially quenched'' theory, with dynamical
quark loops, but, possibly, $m_{\mathrm{valence}}\neq m_{\mathrm{sea}}$.
Then features like the $\eta'$-$\pi$ splitting emerge, and it is
possible to relate the partially quenched theory to full QCD using
chiral perturbation theory.
We shall return to chiral logs in the next section, when
discussing $B^0$-$\bar{B}^0$ mixing.

To understand lattice spacing effects Symanzik suggested
matching lattice gauge theory to continuum
QCD~\cite{Symanzik:1983dc}:
\begin{equation}
	{\cal L}_{\rm lat} \doteq  {\cal L}_{\rm QCD} +
		\sum_i a^{s_i} K_i(g^2,m_qa; \mu) {\cal O}_i(\mu).
\end{equation}
The right-hand side is a local effective Lagrangian (LE${\cal L}$)
renormalized in some continuum scheme; details of the scheme are not
important.
Discretization effects are, of course, short-distance effects, so
as usual in an effective field theory, they
are lumped into the coefficients~$K_i$.

There are two key points to the Symanzik effective field theory.
First, if $\Lambda a$ is small enough, the $\sum_i$
can be treated as a perturbation.
For example, for the proton mass
\begin{equation}
	m_p(a) = m_p - a K_{\sigma F}(c_{\rm SW})
		\langle p|\bar{\psi}\sigma\cdot F\psi|p\rangle
	\label{eq:Sym}
\end{equation}
using the leading term for Wilson fermions as an example.
Here $c_{\rm SW}$ is the so-called clover coupling, and
$K_{\sigma F}=(1-c_{\rm SW})/4+O(\alpha_s)$.
Second, to reduce lattice spacing effects, one can tune $c_{\rm SW}$
so that $K_{\sigma F}$ vanishes or, in practice, is $O(\alpha_s^\ell)$
or~$O(a)$.
Thus, the Symanzik effective field theory shows that lattice artifacts
can be reduced through short-distance process-independent methods.
Indeed, for light hadrons a combination of this Symanzik improvement and
extrapolation in $a^2$ gives continuum QCD results with very well
understood uncertainties (modulo quenching).

With the bottom and charmed quarks, the mass is large in lattice units:
$m_ba\sim$1--2 and $m_{ch}a$ about a third of that.
The split in the LE${\cal L}$ between QCD and small correction breaks 
down, because there are terms in the $\sum_i$ in Eq.~(\ref{eq:Sym}) 
that go like~$(m_Qa)^n$.
It will not be possible to reduce $a$ enough to make $m_ba\ll1$
for a long time: Moore's Law suggests 15--25 years.
There are, nevertheless, several ways to treat heavy-light hadrons in
lattice calculations, all of which appeal to HQET in some way.
Table~\ref{tab:hqlgt} lists the most widely used methods.
\begin{table}
	\begin{tabular}{rlcl}
	\hline
	\# & method & Ref. & how HQET enters \\
	\hline
	1. & static approximation & \cite{Eichten:1987xu,Eichten:1990zv} &
		${\cal L}_{\rm static} = - \bar{h}D_4h$ \\
	2. & lattice NRQCD        & \cite{Lepage:1987gg,Thacker:1991bm} &
		discretization of first few terms of HQET \\
	3. & extrapolation from below charm & \emph{ad hoc}   &
		results with small $m_Q$ fit to Taylor series in $1/m_Q$ \\
	$3'\!.$ & $3+1$ & \emph{ad hoc}   &
		as in method~3 \\
	4. & ``Fermilab''         & \cite{El-Khadra:1997mp,Kronfeld:2000ck} &
		match lattice to QCD, term-by-term in HQET \\
	\hline
	\end{tabular}
	\caption{Widely used methods for heavy quarks in lattice gauge
	theory.}
	\label{tab:hqlgt}
\end{table}
In lattice NRQCD and in the Fermilab method, it is possible to set
$m_Q=m_b$, even when $a^{-1}\sim m_b$.
There are, of course, uncertainties involved, but a grasp of the basics
of heavy-quark theory is enough understand them.

A convenient way to contrast the uncertainties in the various methods is
to match lattice gauge theory to (continuum) HQET~\cite{Kronfeld:2000ck}.
Instead of Eq.~(\ref{eq:Sym}) one writes
\begin{equation}
	{\cal L}_{\rm lat} \doteq {\cal L}_{\rm HQET} = \sum_n
		{\cal C}_n^{\mathrm{lat}} {\cal O}_n =
		- m_1 \bar{h}h - \bar{h}D_4h
		+ \frac{\bar{h}\bbox{D}^2h}{2m_2}
		+ \frac{\bar{h}i\bbox{\Sigma}\cdot\bbox{B}h}{2m_{\cal B}}
		+ \cdots.
\end{equation}
The operators ${\cal O}_n$ on the right hand side are the same as
in the HQET description of continuum~QCD.
The short-distance coefficients~${\cal C}_n^{\mathrm{lat}}$, on the
other hand, are not the same, because there are two short distances,
$a$ and~$1/m_Q$.
Heavy-quark lattice artifacts are, thus, isolated into the mismatch
${\cal C}_n^{\mathrm{lat}}-{\cal C}_n^{\mathrm{cont}}$.
In lattice NRQCD and in the Fermilab method, these mismatches can be
reduced, along the same lines as reducing $K_{\sigma F}$ in the 
Symanzik program.
In recent work, they are controlled to several percent or less.

Most work with method~3 has chosen normalization conditions for which
the mismatch in the kinetic and chromomagnetic terms, although formally
$O(m_Qa)^2$, is quite large in practice.
It is also not well understood how this mismatch is amplified when
extrapolating linearly or quadratically in $1/m_Q$ from $m_Q\sim1$~GeV
up to~$m_b$.

\section{Lattice Calculations}

We now turn our attention to some of the most interesting recent
lattice calculations in $B$ physics.
The discussion focuses on the error bars, and central values are
deferred to the next section.
We consider the ratio $\xi$ for $B$-$\bar{B}$ mixing in unquenched
QCD, and the semi-leptonic decays $B\to\pi l\nu$ and $B\to D^*l\nu$
for $|V_{ub}|$ and $|V_{cb}|$ in quenched QCD.

\subsection{Neutral Meson Mixing}

The mass difference of neutral $B$ meson $CP$ eigenstates is
\begin{equation}
	\Delta m_{B^0_q} = \frac{G_F^2m_W^2S_0}{16\pi^2} |V_{tq}^*V_{tb}|^2
		\eta_B \langle \bar{B}_q^0|Q_q^{\Delta B=2}|B_q^0\rangle
		+ \textrm{new physics?}
\end{equation}
where $q\in\{d,\,s\}$, $S_0$ is an Inami-Lim function, 
$\eta_B$ is a short-distance QCD correction, and
$Q_q^{\Delta B=2}$ is a four-quark operator in the 
$\Delta B=2$ electroweak Hamiltonian.
Note that new physics could compete with the Standard Model.
The matrix element of $Q_q^{\Delta B=2}$ is usually written
\begin{equation}
	\langle \bar{B}_q^0| Q_q^{\Delta B=2}|B_q^0\rangle =
		\frac{8}{3}m^2_{B_q}f^2_{B_q} B_{B_q},
\end{equation}
but lattice QCD gives 
$\langle \bar{B}_q^0| Q_q^{\Delta B=2}|B_q^0\rangle$
and $f_{B_q}$ (from 
$\langle 0| A^\mu_{bd}|B_q^0\rangle$) individually.
This traditional separation turns out to be useful when looking at
chiral logs.

\begin{figure}[b]
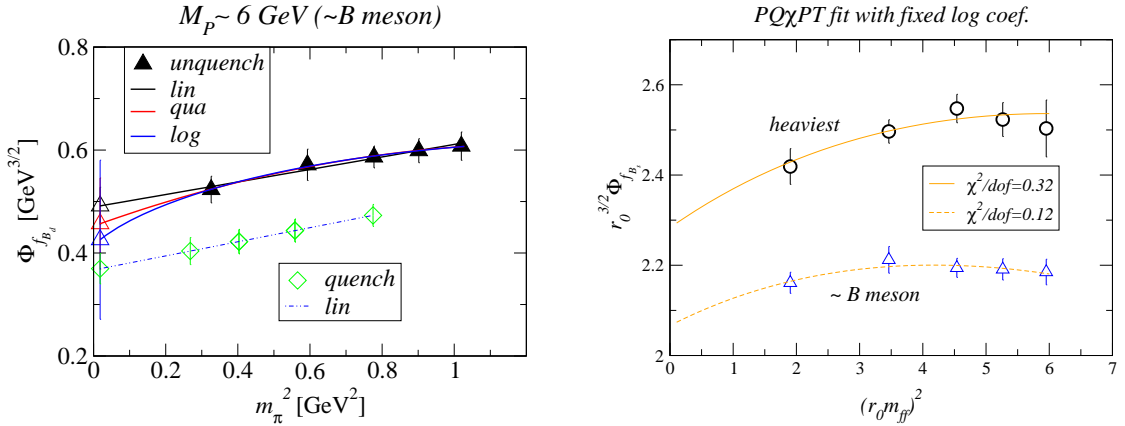

	\includegraphics[width=0.47\textwidth]{naive.eps} \hspace*{0.7cm}
	\includegraphics[width=0.47\textwidth]{chiral.eps}
	\caption{Recent calculations of the decay constant
	$\Phi_B=f_B\sqrt{m_B}$ from JLQCD.}
	\label{fig:jlqcd}
\end{figure}

Conventional wisdom says that the uncertainties in $B_{B_q}$ and in
$\xi^2=f_{B_s}^2B_{B_s}/f_{B_d}^2B_{B_d}$ are ``easy'' to control,
because they are ratios.
At {\sl Lattice~'01}, Norikazu Yamada of JLQCD reported on new
results~\cite{Yamada:2001xp} that suggest otherwise.
The JLQCD collaboration is mounting a large project to carry out
calculation with the loops of $n_f=2$ flavors of quarks.
Figure~\ref{fig:jlqcd} compares their previous quenched work for
$f_{B_q}$ with their new (and still preliminary) work with $n_f=2$.
In the spirit of a blind analysis, the definition of $\Phi_{f_{Bq}}$ is
not so important.
The important matter is whether the curve is a straight line, or whether
there is any curvature as a function of the light quark mass.
(Because the pseudoscalar $m_\pi^2\propto m_q$, the horizontal axis is,
essentially, the light quark mass.)

The quenched approximation (diamonds in Fig.~\ref{fig:jlqcd})
shows no evidence of curvature.
With $n_f=2$ (triangles in Fig.~\ref{fig:jlqcd}), however, allowing 
for curvature obtains a better fit.  In fact, curvature is expected 
from chiral logarithms.
Meson loops with fixed spectator mass $m_s$ and varying sea quark 
mass~$m_f$ contribute to $f_{B_s}$~\cite{Sharpe:1996qp}
\begin{equation}
	\Delta f_{B_s} = c_0 + c_1 m^2_{\eta_{ff}} - 
		\frac{1+3g^2}{(4\pi f_\pi)^2} m^2_{\eta_{sf}}
		\ln\left( \frac{m^2_{\eta_{sf}}}{m^2_{\eta_{ss}}} \right),
	\label{eq:chlog}
\end{equation}
where $\eta_{qq'}$ denotes a pseudoscalar meson with constituents $q$
and~$q'$.
From the $D^*$ width, $3g^2\approx 1.04$~\cite{Anastassov:2001cw}.
The second plot in Fig.~\ref{fig:jlqcd} shows that Eq.~(\ref{eq:chlog})
describes the points well.

It is worth making a few remarks about this kind of analysis.
The formula for $f_{B_d}$ (where the spectator and sea quarks have
the same mass) is slightly different, so it is difficult to display
all chiral log effects on one plot.
Nevertheless, one can see from Fig.~\ref{fig:jlqcd} that the effect
is to increase $f_{B_s}/f_{B_d}$.
The chiral logs for $B_{B_q}$ are multiplied by $1-3g^2\approx0.04$,
so the $B_{B_s}/B_{B_d}$ is almost insensitive to them.
Therefore, the chiral logs in the ratio~$\xi$ come almost completely
from $f_{B_s}/f_{B_d}$. 
One concludes that $\xi$ may have been underestimated in the quenched 
approximation.
A~new estimate, based on JLQCD and previous work, is given below.

\subsection{$B\to\pi l\nu$ and $V_{ub}$}

The semi-leptonic decay $B\to\pi l\nu$ is mediated by a $b\to u$
transition.
The decay rate is
\begin{equation}
	\frac{d\Gamma}{dE} = \frac{G_F^2m_Bp^3}{12\pi^3}
		|V_{ub}|^2 |f_+(E)|^2,
\end{equation}
where $v=p_B/m_B$ is the $B$ meson's velocity,
$E=v\cdot p_\pi$, and $p^2=E^2-m_\pi^2$.
The form factor $f_+(E)$ is a linear combination of the form factors
$f_\perp(E)$ and $f_\parallel(E)$, defined through the matrix element
of the vector current
\begin{equation}
	\langle\pi|V^\mu|B\rangle = \sqrt{2m_B}
		\left[v^\mu f_\parallel(E) + p^\mu_\perp f_\perp(E)\right].
\end{equation}
The pion energy~$E$ is related to $q^2=m_B^2+m_\pi^2-2m_BE$.
Chiral symmetry and heavy-quark symmetry are simpler to follow with 
$f_\parallel$ and $f_\perp$ considered as functions of~$E$, rather 
than $f_+$ considered as a function of $q^2$.
For example, heavy-quark symmetry suggests relations between $B$ and 
$D$ decays with the same~$E$.
In principle, one would like to compare the $E$ dependence of
experimental measurements with lattice calculations.
Discretization uncertainties are smallest for low~$E$, where phase 
space suppresses the event rate.
Therefore, lattice calculations and experimental measurements will
have to be combined in the way that minimizes the error on~$|V_{ub}|$.

In the past year or so, four new quenched calculations of
$f_+(E)$ for $B\to\pi l\nu$ appeared, using several different
methods~\cite{Bowler:1999xn,Abada:2000ty,El-Khadra:2001rv,Aoki:2001rd}.
It is, thus, timely to compare and contrast.
In addition to discretization effects at large~$pa$, there is
evidence for considerable dependence on the light spectator quark
mass~\cite{El-Khadra:2001rv}.

The calculations of UKQCD~\cite{Bowler:1999xn} and
APE~\cite{Abada:2000ty} use method~3, so they have
\linebreak
\begin{minipage}{0.52\textwidth}
discretization errors of order
$(m_Qa)^2$.
They keep $m_Q<1.2m_c$ and extrapolate up to $m_b$ linearly or
quadratically in $1/m_Q$.
The quark masses used in Ref.~\cite{Bowler:1999xn} are shown in
Table~2.
\addtocounter{table}{1}
Those in Ref.~\cite{Abada:2000ty} are a bit larger, which is better for
HQET but worse for the discretization errors.
In these papers models are used to extrapolate to the full kinematic
range of~$E$.
It is better to think of these calculations as computing the model
parameters in $D$ decays, and then invoking HQET to make predictions
for \linebreak
\end{minipage} \hfill
\begin{minipage}{0.44\textwidth}
%\begin{table}[h]
%	\caption
	{\fontsize{10}{11.5}\bfseries TABLE~2.}
	{\fontsize{10}{11.5}\selectfont 
	Heavy quark masses used in \linebreak Ref.~\cite{Bowler:1999xn}.
	Note that $m_{\textrm{pole}} < m^{\rm RI}$.}

	\begin{tabular}{cccc}
\hline
		$\kappa$ & $am_Q$ & ${m}^{\rm RI}_Q$ & $M_P$ \\
				 &        & (GeV) & (GeV)\\
	\hline
		0.1200  & 0.485   & 1.52 & $2.035(5)$\\
		0.1233  & 0.374   & 1.23 & $1.771(5)$ \\
		0.1266  & 0.268   & 1.02 & $1.483(5)$\\
		0.1299  & 0.168   & 0.69 & $1.157(5)$\\
\hline
	\end{tabular}
%\end{table}
\end{minipage}\\[-10pt]
$B$ decays.
The error associated with the assumptions are, at least to me,
not transparent.

\begin{figure}[b]
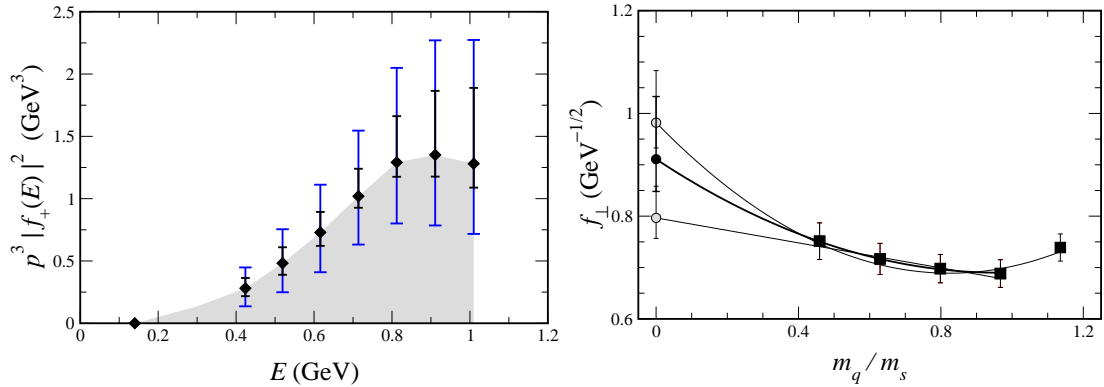

	\includegraphics[height=2.0in]{B_dGamma_dp_sys.eps}
	\includegraphics[height=2.0in]{chiral_fperp_p100.eps}
	\caption{$B\to\pi l\nu$: (a) Integrand of Eq.~(\ref{eq:TB}).
	(b) Dependence of the form factor $f_\perp$ on the light quark 
	mass, in the range $0.4m_s<m_q<1.2m_s$.}
	\label{fig:fperp}
\end{figure}

The calculation of El-Khadra {\em et al.}~\cite{El-Khadra:2001rv} uses
the Fermilab method, and the calculation of JLQCD~\cite{Aoki:2001rd}
uses lattice NRQCD.
Since the matching procedures build in the heavy-quark expansion,
both calculate directly at $m_Q=m_b$.
To avoid introducing models of the $E$ dependence, these two papers
advocate comparing lattice and experiment in a region where the two
overlap.
For example, one can look at
\begin{equation}
	T_B(E_{\mathrm{max}}) =
		\int_{m_\pi}^{E_{\mathrm{max}}} \!\!dE\; p^3 |f_+(E)|^2 =
		\frac{1}{|V_{ub}|^2}
		\frac{12\pi^3}{G_F^2m_B}
		\int_{m_\pi}^{E_{\mathrm{max}}} \!\!dE\; \frac{d\Gamma}{dE},
	\label{eq:TB}
\end{equation}
where $E_{\mathrm{max}}$ is an upper kinematic cut.
At present $E_{\mathrm{max}}\sim 1$~GeV, but, with future increases 
in computer power, it may be possible to raise the cut.
This method is sketched in Fig.~\ref{fig:fperp} using the form factor
from Ref.~\cite{El-Khadra:2001rv}.
Uncertainties are still about 15--20\% on~$|V_{ub}|$.
A~strategy for reducing them are given in Ref.~\cite{El-Khadra:2001rv}.

Ref.~\cite{El-Khadra:2001rv} finds strong dependence on the light 
quark mass, as illustrated for $f_\perp(0.7~\mathrm{GeV})$ in 
Fig.~\ref{fig:fperp}.
This is the only calculation to reduce the light quark down to
$m_q=0.4m_s$.
The extrapolation to physical quark mass is the largest source of
uncertainty quoted by Ref.~\cite{El-Khadra:2001rv}, and a similar 
uncertainty is presumably present in all four calculations.
One would like even lighter quarks, which take more CPU time.

Because of the discretization errors at higher~$E$, the large light
quark mass dependence, etc., a sanity check on $f_+(E)$ for 
$B\to\pi l\nu$ would be welcome.
The calculation of similar form factors in $D$ decays encounters many 
of the same issues.
A compelling test would be to compare lattice and experiment for
$f_+^{D\to K}/f_{D_s}$ and $f_+^{D\to\pi}/f_D$, as a function of~$E$.
New physics is unlikely to alter the underlying processes, and
the CKM matrix drops out.
The CLEO-c program~\cite{Briere:2001rn} promises to measure these 
ratios with an uncertainty of a few percent.

\subsection{$B\to D^*l\nu$ and $V_{cb}$}

The exclusive semi-leptonic decay $B\to D^*l\nu$ is a good way to
determine~$|V_{cb}|$, but an estimate of the hadronic transition
is needed.
The differential decay rate is
\begin{equation}
	\frac{d\Gamma}{dw} = \frac{G_F^2}{4\pi^3}\sqrt{w^2-1}
		m_{D^*}^3 (m_B-m_{D^*})^2 {\cal G}(w)
		|V_{cb}|^2 |{\cal F}_{B\to D^*}(w)|^2,
\end{equation}
where $w=v\cdot v'$ and ${\cal G}(1)=1$.
At zero recoil ($w=1$) heavy-quark flavor symmetry forbids terms of 
order $1/m_Q$.
The exclusive determination of $|V_{cb}|$ therefore proceeds by 
extracting $|V_{cb}{\cal F}_{B\to D^*}(1)|$ from a fit to the 
measurement of $d\Gamma/dw$, and then taking a theoretical estimate of 
${\cal F}_{B\to D^*}(1)$.
One would prefer the latter not to depend on models.

To see what is needed from the theoretical calculation, let us review 
the anatomy of ${\cal F}_{B\to D^*}(1)$.
From HQET one can show that, at zero recoil, 
\begin{equation}
	{\cal F}_{B\to D^*}(1) = \eta_A
		\left[1_{\mathrm{Isgur-Wise}} + 0_{\mathrm{Luke}}/m +
		\delta_{1/m^2} + \delta_{1/m^3} \right]
	\label{eq:F}
\end{equation}
where $\eta_A$ is a short-distance radiative correction, and the
$\delta_{1/m^n}$ contain the long-distance properties of the bound 
states.
The numerical values of $\eta_A$ and the $\delta_{1/m^n}$ depend on 
how HQET is renormalized.
It is less important \emph{which} renormalization scheme is chosen 
than it is to use the \emph{same} scheme for both.
The $1/m_Q^n$ corrections take the form
\begin{eqnarray}
	\delta_{1/m^2} & = & - \frac{\ell_V}{(2m_c)^2}
		+ \frac{2\ell_A}{(2m_c)(2m_b)} - \frac{\ell_P}{(2m_b)^2} ,
	\label{eq:delta2} \\
	\delta_{1/m^3} & = &
		- \frac{\ell_V^{(3)}}{(2m_c)^3}
		+ \frac{\ell_A^{(3)}\Sigma + \ell_D^{(3)}\Delta}{(2m_c)(2m_b)}
		- \frac{\ell_P^{(3)}}{(2m_b)^3} ,
	\label{eq:delta3}
\end{eqnarray}
where $\Sigma=1/(2m_c)+1/(2m_b)$ and $\Delta=1/(2m_c)-1/(2m_b)$.

With brute force alone, a sufficiently precise calculation of 
${\cal F}_{B\to D^*}(1)$ lies beyond reach.
(See Ref.~\cite{Hashimoto:2001nb} for details.)
Hashimoto \emph{et al.}~\cite{Hashimoto:2001nb} have devised a method to
extract all $\ell$s in Eq.~(\ref{eq:delta2}) and all but $\ell_D^{(3)}$
in Eq.~(\ref{eq:delta3}).
The key is the heavy-quark mass dependence of three double ratios of
matrix elements.
As in Eqs.~(\ref{eq:F})--(\ref{eq:delta3}), HQET implies
\begin{eqnarray}
	\frac{	\langle D   |\bar{c}\gamma^4 b| B   \rangle
			\langle B   |\bar{b}\gamma^4 c| D   \rangle}
		{	\langle D   |\bar{c}\gamma^4 c| D   \rangle
			\langle B   |\bar{b}\gamma^4 b| B   \rangle} & = & \eta_V^2
	\left[1 - \Delta^2\left(\ell_P + \ell^{(3)}_P\Sigma \right)\right]^2,
	\label{eq:R+} \\
	\frac{	\langle D^* |\bar{c}\gamma^4 b| B^* \rangle
			\langle B^* |\bar{b}\gamma^4 c| D^* \rangle}
		{	\langle D^* |\bar{c}\gamma^4 c| D^* \rangle
			\langle B^* |\bar{b}\gamma^4 b| B^* \rangle} & = & \eta_V^2
	\left[1 - \Delta^2\left(\ell_V + \ell^{(3)}_V\Sigma \right)\right]^2,
	\label{eq:R1} \\
	\frac{	\langle D^* |\bar{c}\gamma^j \gamma_5 b| B   \rangle
			\langle B^* |\bar{b}\gamma^j \gamma_5 c| D   \rangle}
		{	\langle D^* |\bar{c}\gamma^j \gamma_5 c| D   \rangle
			\langle B^* |\bar{b}\gamma^j \gamma_5 b| B   \rangle} & = &
	\frac{\eta_{A^{cb}}\eta_{A^{bc}}}{\eta_{A^{cc}}\eta_{A^{bb}}}
	\left[1 - \Delta^2\left(\ell_A + \ell^{(3)}_A\Sigma \right)\right]^2.
	\label{eq:RA1}
\end{eqnarray}
In particular, heavy-quark spin symmetry requires the same $\ell$s to 
appear in Eqs.~(\ref{eq:R+})--(\ref{eq:RA1}) as in 
Eqs.~(\ref{eq:delta2}) and~(\ref{eq:delta3}).

In a lattice calculation of these double ratios most of the 
statistical and sytematic uncertainties cancel.
The main difference from continuum QCD is that the short-distance 
coefficients are different~\cite{Kronfeld:2000ck,Hashimoto:2001nb}.
But, just like in continuum QCD, the short-distance behavior can be 
computed in perturbation theory.
Thus, the analysis proceeds by removing the lattice short-distance
contribution, fitting to the prediction of HQET, and then reconstituting
$\mathcal{F}_{B\to D^*}(1)$ from the $\ell$s and~$\eta_A$.
The result is
\begin{equation}
	\mathcal{F}_{B\to D^*}(1) = 0.9??^{+0.024}_{-0.017} \pm 0.016
		{}^{+0.003}_{-0.014} {}^{+0.000}_{-0.016} {}^{+0.006}_{-0.014}.
	\label{eq:Fblind}
\end{equation}
To encourage the reader to think about the uncertainties, the central 
value is not revealed until the end.
The error bars come from, in order,
statistics and fitting,
matching lattice gauge theory with HQET to QCD,
lattice spacing dependence,
light quark mass effects,
\emph{and} the quenched approximation.
The secret to the small error on ${\cal F}$ is that they all scale as 
${\cal F}-1$, by design.
As a fraction of ${\cal F}-1$ the uncertainties are still sizable: 
5--25\%.

The most novel aspect of this analysis is how seriously it takes the 
idea of matching lattice gauge theory to QCD through HQET.
A~central part of the analysis is to calculate short-distance 
properties, which Ref.~\cite{Hashimoto:2001nb} does partly in 
perturbation theory.
The second error bar reflects the associated uncertainty.
It is reducible, but through traditional theoretical physics, 
rather than intensive computation.

\section{An Octopus's Garden}

Although it does not bear directly the unitarity triangle, I would 
like to discuss some recent work by Nathan Isgur.
I did not know Nathan well, but his enthusiasm for physics, 
especially the strong interactions, made a big impression on~me.
He struck me as the kind of person who had a hand all sorts of things.
He must have needed eight arms to keep all his projects moving.

In the last year of his life, one of his projects was to understand 
whether instanton-like gauge fields or, on the other hand, disordered 
gauge fields play a larger role
in~QCD~\cite{Isgur:2000ts,Horvath:2001ir,Bardeen:2001jm}.
This work reflected Nathan's broad knowledge of the phenomenological 
and theoretical sides of the strong interactions, touching on the OZI 
rule, chiral zero modes, the $\eta'$ mass, instantons, the quark 
model, the AdS/CFT correspondence, lattice QCD, and the large $N_c$ 
limit.
A particularly striking passage notes first how recent work by Witten 
on the AdS/CFT correspondance favors disordered, confining gauge 
fields (although those sentences were probably written by Nathan's 
collaborator Hank Thacker), and then how details of how to treat 
strong decays in the quark model favor the disorder also.
That part was \emph{certainly} written by Nathan.

These issues surround the quantum ground state, or vacuum, of~QCD.
Gauge theories have many classical ground states, one for each integer.
Instantons are the semi-classical configurations that tunnel from one 
ground state to another.
In a quantum mechanical situation, fluctuation-dominated 
configurations could also mediate tunneling.
Nathan proceeded by devising tests that could distinguish whether the 
quantum vacuum is obtained via such disordered gauge fields or via a 
gas or liquid of instantons.
He then used (quenched) lattice QCD to see which way the gauge fields 
behaved.

One test stemmed from the observation, from empirically successful
quark models of strong decays, that quark pairs pop out of the vacuum
with scalar (\emph{i.e.}, $^3P_0$) quantum numbers.
That implies that the OZI rule should fail for $J^{PC}=0^{++}$ 
quantum numbers.
(The OZI rule says that decays, like $\phi\to\pi\pi$, in which quarks 
annihilate, are weaker than those, like $\phi\to KK$, in which they 
do not.)
Empirically, the OZI rule fails in the $0^{-+}$ sector; that is how 
the $\eta'$ mass is split away from the other pseudoscalar mesons.
For the $0^{++}$ sector, however, there is a dearth of experimental 
data.
Isgur and Thacker's idea to to compute the quark annihilation process
with lattice QCD, for various $J^{PC}$, which is possible at leading
order in the OZI approximation even in quenched QCD~\cite{Isgur:2000ts}.
They found a small amplitude where the OZI rule holds empirically,
\emph{e.g.}, for $J^{PC}=1^{-+}$.
But for scalar and pseudoscalar channels the amplitude is large.
This is evidence against an instanton-dominated vacuum, because 
instantons couple to quarks through the axial anomaly, that is, 
preferentially to the pseudoscalar channel.

\section{Unblinding}

Instead of a paragraph of bland conclusions, the reward for readers 
who have made it this far consists of the most interesting numerical 
results.
The preliminary results of JLQCD on $B$ physics with $n_f=2$
are~\cite{Yamada:2001xp}
\begin{eqnarray}
	f_{B_d}         & = & 190(14)(7)(19)~{\rm MeV}, \\
	f_{B_s}/f_{B_d} & = & 1.184(26)(20)(15), \\
	\xi             & = & 1.183(27)(20)(15),
\end{eqnarray}
where $\xi$ is a combination of their results for
$f_{B_s}/f_{B_d}$ and $B_{B_s}/B_{B_d}$.
In her review at {\sl Lattice~'01}~\cite{Ryan:2001ej}, Sin\'ead Ryan
took stock of the ratio~$\xi$, which enters into UT fits.
Although JLQCD's result is still preliminary, it cannot be denied that
the chiral logarithms could affect $f_{B_s}/f_{B_d}$ and, hence, $\xi$.
Ryan's average, which strikes me as reasonable, is
\begin{equation}
	\xi = 1.15 \pm 0.06 ^{+0.07}_{-0.00}
\end{equation}
which retains the central value in common usage, but allows for a
future upward revision, once the chiral behavior is fully understood
and controlled.

For the zero-recoil form factor needed to determine $|V_{cb}|$ from
$B\to D^*l\nu$, Ref.~\cite{Hashimoto:2001nb} finds
\begin{equation}
	{\cal F}_{B\to D^*}(1) = 0.913^{+0.024}_{-0.017}{}^{+0.017}_{-0.030},
\end{equation}
with systematics added in quadrature.
Note that this result does \emph{not} include the QED correction of
+0.007, which is included, for example, in the {\sl BaBar} Physics Book.
This result still relies on the quenched approximation.
Nevertheless, it is probably still under better control than 
estimates based on the quark model. 

A hallmark of these two results, shared by some work on $B\to\pi l\nu$
for $|V_{ub}|$~\cite{El-Khadra:2001rv,Aoki:2001rd}, is that they attempt
a full analysis of the uncertainties, including those of heavy quarks,
the chiral behavior, and quenching.
Thus, they are suitable templates for the next round of lattice 
calculations, from which one can expect serious \emph{unquenched} 
calculations with a direct impact on our knowledge of the unitarity 
triangle.

\begin{theacknowledgments}
It is a pleasure to thank Shoji Hashimoto, Hank Thacker, and Norikazu
Yamada for helpful correspondance while preparing this talk.
The organizers and the staff at CalTech managed the symposium superbly,
at a time when we were all distracted and appalled by world events.
Fermilab is operated by Universities Research Association Inc.,
under contract with the U.S.\ Department of Energy.
\end{theacknowledgments}

\end{document}